\documentclass{jfm}
\usepackage{graphicx}
\usepackage{epstopdf, epsfig}
\usepackage{tikz}
\usetikzlibrary{arrows,shapes}
\usepackage{caption}
\usepackage{subcaption}

\shorttitle{Nonlinear forcing in ECS}
\shortauthor{K. Rosenberg and B.J. McKeon}

\title{Nonlinear forcing in the resolvent analysis of exact coherent states of the Navier-Stokes equations}

\author{Kevin Rosenberg\aff{1}
  \corresp{\email{krosenbe@caltech.edu}}
 \and Beverley J. McKeon\aff{1}}

\affiliation{\aff{1}Graduate Aerospace Laboratories, California Institute of Technology,
Pasadena, CA 91125, USA}

\begin{document}

\maketitle
% % % % % % % % % % % % % % % % % ABSTRACT % % % % % % % % % % % % % % % % % % % % % % % % % % % % % %
\begin{abstract}
The resolvent analysis of \citet{mckeon2010critical} recasts the Navier-Stokes equations into an input/output form in which the nonlinear term is treated as a forcing that acts upon the linear dynamics to yield a velocity response. The framework has shown
promise with regards to producing low-dimensional representations of exact coherent
states. Previous work has focused on a primitive variable output; here we show a velocity-vorticity formulation of the governing equations along with a Helmholtz decomposition of the nonlinear forcing term reveals a simplified input/output form in the resolvent analysis. This approach leads to an improved method for compact representations of exact coherent states for both forcing and response fields, with a significant reduction in degrees of freedom in comparison to the primitive variable approach.
\end{abstract}

\begin{keywords}
resolvent, nonlinear forcing, exact coherent states
\end{keywords}
% % % % % % % % % % % % % % % % % INTRODUCTION % % % % % % % % % % % % % % % % % % % % % % % % % % % % % %
\section{Introduction}
\subsection{Background}
Evidence of coherent structures in turbulent flows has motivated the development of low-order models to describe the underlying dynamics. There are many existing approaches and they can be broadly split into two categories: data-driven, such as proper orthogonal decomposition and dynamic mode decomposition, or operator-driven, such as Koopman mode decomposition and resolvent analysis. In both instances, the aim is to construct an optimal basis (in some defined sense) to represent the flow field; the former relies on snapshots of data from experiments or simulations while the latter exploits the mathematical structure of the governing equations. An extensive overview of existing techniques and their use in the literature is found in \citet{rowley2017model}. One of the  ultimate goals is to use these low-order models as a means to gain insight into the fundamental physical processes of the flows of interest and leverage this knowledge to develop control strategies to achieve some desired effect, such as turbulent drag reduction.

Herein, we focus on the resolvent model of \citet{mckeon2010critical} for turbulent flows. This approach re-formulates the Navier-Stokes equations (NSE) into an input/output form where the nonlinear term, treated as a forcing, acts on the linear dynamics to yield a response across wavenumber/frequency space. In general, the response can be any linear combination of the states of the system, though much of the work to the date has focused on primitive variables. The linear operator which maps input to output is known as the resolvent operator, and will be discussed in more detail in \S\ref{sec:RA}. The resolvent contains important information regarding the linear mechanisms by which fluctuations (modes) become amplified and organized via interactions with the mean flow. Notable mechanisms include the critical layer, the location where the wavespeed of the mode matches the local mean velocity, and lift-up effects due to interactions with the mean shear. A singular value decomposition (SVD) of the resolvent operator is used to identify highly amplified input/output pairs as a means of obtaining low-dimensional representations of flow fields. A summary of the progress of the resolvent model for analyzing turbulent flows is found in \citet{mckeon_2017} . While the resolvent operator has been well studied for canonical turbulent flows including pipes and channels, the nonlinear forcing (i.e. the interaction of modes) that drives the resolvent has received less attention. Gaining insight into the role of the forcing term is crucial to the understanding of self-sustaining mechanisms, and their study remains difficult for fully developed turbulent flows due to the wide range of scales and the limited availability of data in the full wavenumber/frequency domain. Recently, \citet{sharma2016low} investigated the use of resolvent analysis for low-dimensional representations of exact coherent states of the NSE, due partly to their decreased complexity; this work seeks to expand upon this previous study with a more in-depth analysis of the forcing term and its potential for improved compact representation. In what follows, \S\ref{sec:ECS} provides some background on exact coherent states and the motivation for analyzing them, \S\ref{sec:RA} formulates the governing relations in the resolvent analysis and the treatment of the forcing term, \S\ref{sec:results} presents results, and \S\ref{sec:conclusions} offers some concluding remarks.
 
\subsection{Exact coherent states}\label{sec:ECS}
Exact coherent states \citep{waleffe2001exact}(ECS) are nonlinear travelling wave solutions to the NSE. ECS, along with other exact solutions (e.g. equilibria and periodic orbits), have been hypothesized to constitute the state-space skeleton of turbulent dynamics. Families of solutions have been found for both pipe and channel geometries. Solutions primarily arise in pairs due to a bifurcation at a finite Reynolds number, where the lower branch (L) solutions have lower drag in comparison to the upper branch (U) solutions. One particular family of solutions for the channel, termed P4, have been linked to trajectories frequently visited by fully-developed turbulence and are thus of practical interest with regards to understanding the underlying dynamics of turbulent flows \citep{park2015exact}. Furthermore, the main energetic activity of the P4 solutions is known to be concentrated around the critical layer, a prominent mechanism in the resolvent framework. The corresponding P4L and P4U solutions will be the focus of analysis in this paper. In addition to their relevance to turbulent dynamics, the solutions present tractable datasets to analyze as they convect at a single streamwise wavespeed, which greatly simplifies the frequency domain analysis, and are often solved for in minimal domains and thus significantly reduce the number of degrees of freedom in the system as compared to direct numerical simulations of fully turbulent flows. The computation of these solutions utilizes the code \textit{channelflow} \citep{gibson2008visualizing}, which employs a Fourier/Chebyshev/Fourier discretization in the streamwise (x), wall-normal (y), and spanwise (z) directions respectively. A summary of the relevant geometrical parameters and flow quantities for the solutions investigated here are found in table 1; here $Re_{\tau}$ is the friction Reynolds number, $L_i$ denotes the domain in the $i^{th}$ spatial direction, $N_i$ denotes the corresponding number of discretization points, and $c^+$ is the wavespeed (normalized in inner units) of the lower and upper branch solutions respectively. It should be noted these solutions have imposed reflection symmetries with respect to the midplanes of the wall-normal and spanwise directions. We focus our attention to a fixed friction Reynolds number for simplicity of presentation. We next derive the mathematical framework behind the resolvent analysis and discuss how it may be used to represent these ECS solutions in an efficient manner.

\begin{table}\label{tab:ECS_param}
  \begin{center}
\def~{\hphantom{0}}
  \begin{tabular}{ccccccccccc}
       $Re_{\tau}$   & $L_x$ & $L_y$ & $L_z$ &   $N_x$ & $N_y$ & $N_z$ & $c_L$ & $c_U$ \\[3pt]
       85  & [0,$\pi$] & [-1,1] & [0,$\pi/2$]&  24 & 81 & 24 & 25 & 14.2\\
  \end{tabular}
  \caption{Traveling wave solution parameters and domain discretization.}
  \label{tab:kd}
  \end{center}
\end{table}
% % % % % % % % % % % % % % % % % RESOLVENT ANALYSIS % % % % % % % % % % % % % % % % % % % % % % % % % % %
\section{Resolvent analysis}\label{sec:RA}
\subsection{Formulation}

We consider the non-dimensional, incompressible NSE for a Newtonian fluid in a channel
\begin{equation}\label{eq:NSE}
\frac{\partial \mathbf{u}}{\partial t} + \mathbf{u} \cdot \nabla \mathbf{u} = -\nabla p + \frac{1}{Re_\tau} \nabla^2 \mathbf{u}
\end{equation}
\begin{equation}\label{eq:Continuity}
\nabla \cdot \mathbf{u} = 0
\end{equation}
where ${Re_\tau} = \frac{hu_{\tau}}{\nu}$, h is the channel half-height, $u_{\tau} = \sqrt{\frac{\tau_w}{\rho}}$ is the friction velocity, $\tau_{w}$ is the wall shear stress, $\rho$ is the density, and $\nu$ is the kinematic viscosity. The velocity is non-dimensionalized by $u_{\tau}$, time by $\frac{h}{u_{\tau}}$, spatial variables by $h$, and pressure by $ \rho {u_{\tau}}^2$. The streamwise direction and spanwise directions are periodic, and the wall-normal domain extends from $y/h =-1$ to $y/h=1$ with no-slip and no-penetration conditions imposed at the wall. The velocity field is Reynolds decomposed into the sum of a spatio-temporal mean and fluctuations,
\begin{equation}\label{eq:RD}
\mathbf{u}(x,y,z,t) = \mathbf{U}(y) + \mathbf{u'}(x,y,z,t).
\end{equation}
The fluctuations are most efficiently expressed as Fourier modes in the homogeneous directions,
\begin{equation}\label{eq:FT}
\hat{\mathbf{u}}(k_x,k_z,\omega;y) = \displaystyle\int_{-\infty}^\infty \displaystyle\int_{-\infty}^\infty \displaystyle\int_{-\infty}^\infty \mathbf{u}'(x,y,z,t)e^{-i(k_xx + k_zz -\omega t)}\mathrm{d}x \mathrm{d}z \mathrm{d}t
\end{equation}
where $k_x$ is the streamwise wavenumber, $ k_z $ is the spanwise wavenumber, and $\omega$ is the radial frequency. Substitution of \ref{eq:RD} - \ref{eq:FT} into \ref{eq:NSE} and elimination of the pressure term yields the following set of equations in terms of the fluctuating vertical velocity $ \hat{v} $ and normal vorticity $ \hat{\eta} = ik_z \hat{u} - ik_x \hat{w}$,
\begin{equation}\label{eq:OS_SQ}
-i\omega 
\left(
\begin{array}{c}
\hat{v} \\ \hat{\eta} 
\end{array}\right)
+\left(
\begin{array}{c c}
\Delta^{-1} & 0 \\ 0  & 1  \end{array} \right)
\left(
\begin{array}{c c}
\mathcal{L}_{OS} & 0 \\ ik_zU'  & \mathcal{L}_{SQ}
\end{array}\right)
\left(
\begin{array}{c}
\hat{v} \\ \hat{\eta} 
\end{array}\right)
= B\hat{\mathbf{f}}
\end{equation}
where the Orr-Sommerfeld and Squire operators are given by
\begin{eqnarray}
\mathcal{L}_{OS} = ik_xU\Delta - ik_xU'' -\frac{1}{Re_{\tau}}\Delta^2 \\
\mathcal{L}_{SQ} = ik_xU - \frac{1}{Re_{\tau}}\Delta,
\end{eqnarray}
and the forcing operator is
\begin{equation}\label{eq:B_op}
B = \left(
\begin{array}{c c}
\Delta^{-1} & 0 \\ 0  & 1 
\end{array}\right)
\left(
\begin{array}{c c c}
-ik_x \frac{\partial}{\partial{y}} & -k^2 & -ik_z \frac{\partial}{\partial{y}} \\
  ik_z & 0 & -ik_x 
\end{array}\right),
\end{equation}
\begin{equation}
\hat{\mathbf{f}}=
\left(
\begin{array}{c}
\hat{f}_u \\ \hat{f}_v \\ \hat{f}_w  
\end{array}\right) = -\langle \mathbf{u}' \cdot \nabla \mathbf{u}' \rangle_\mathbf{k}.
\end{equation}
Here $\Delta = \frac{\partial ^2}{\partial y^2} - k^2$, $ k^2 = k_x^2 + k_z^2$, and $\langle \hspace{2 mm} \rangle_{\mathbf{k}} $ denotes the Fourier component associated with the wavenumber vector $ \mathbf{k} = (k_x,k_z,\omega)$. We emphasize that equation \ref{eq:OS_SQ} is an exact form of the NSE and the nonlinearity is explicitly retained in $\hat{\mathbf{f}}$. The equation can be recast into the following input/output form between the forcing $\hat{\mathbf{f}}$ and a general output $\hat{\mathbf{g}}$,
\begin{equation}
\hat{\mathbf{g}} = \mathcal{H}(k_x,k_z,\omega)\hat{\mathbf{f}} 
\end{equation}
where the transfer function $\mathcal{H}$, henceforth referred to as the resolvent operator, is given by
\begin{equation}
\mathcal{H}(k_x,k_z,\omega) = C(-i\omega+A)^{-1}B, 
\end{equation}
\begin{equation}
A = \left(
\begin{array}{c c}
\Delta^{-1} & 0 \\ 0  & 1  \end{array} \right)
\left(
\begin{array}{c c}
\mathcal{L}_{OS} & 0 \\ ik_zU'  & \mathcal{L}_{SQ}
\end{array}\right),
\end{equation}
and $C$ dictates the desired outputs. In the case of a primitive variable formulation,
\begin{equation}\label{eq:PrimitiveResolvent}
\left(
\begin{array}{c}
\hat{u} \\ \hat{v} \\ \hat{w}  
\end{array}\right) = \mathcal{H}_{p}(k_x,k_z,\omega)\left(
\begin{array}{c}
\hat{f}_u \\ \hat{f}_v \\ \hat{f}_w  
\end{array}\right), 
\end{equation}
C is given by
\begin{equation}\label{eq:C_op}
C = \frac{1}{k^2}\left(
\begin{array}{c c c}
ik_x \frac{\partial}{\partial{y}} & -ik_z\\ k^2 & 0\\ ik_z \frac{\partial}{\partial{y}}  & ik_x  \end{array} \right).
\end{equation}
Low order representations of the velocity and forcing fields are obtained by considering a SVD of the resolvent operator,
\begin{equation}
\mathcal{H}_p(k_x,k_z,\omega) = \Psi \Sigma \Phi^T 
\end{equation}
where $\Phi$ contains an ordered set of orthogonal input modes $\left\lbrace\!
\begin{array}{c c c c}\hat{\phi}_{{1}}(y) & \hat{\phi}_{{2}}(y) & \cdots & \hat{\phi}_{{n}}(y) \end{array}\!\right\rbrace $, $\Psi$ contains an ordered set of orthogonal response modes $\left\lbrace\!
\begin{array}{c c c c}\hat{\psi}_{{1}}(y) & \hat{\psi}_{{2}}(y) & \cdots & \hat{\psi}_{{n}}(y) \end{array}\!\right\rbrace $, and $\Sigma$ contains the corresponding singular values $\left\lbrace\!
\begin{array}{c c c c}\sigma_1 & \sigma_2 & \cdots & \sigma_n \end{array}\!\right\rbrace $, i.e. the gains between input/response pairs, where $\sigma_1 \ge \sigma_2 \ge \cdots \sigma_n$. The SVD is optimal with respect to an L2-norm, and thus the response modes in the primitive variable formulation are optimal in the sense of capturing the kinetic energy of a particular Fourier mode. With this decomposition, we can approximate the velocity field as a finite sum of weighted response modes,
\begin{equation}
\hat{\mathbf{u}}(k_x,k_z,\omega;y) \approx \displaystyle\sum_{j=1}^{N} \sigma_j(k_x,k_z,\omega)\chi_j(k_x,k_z,\omega)\hat{\psi}_j(k_x,k_z,\omega;y) 
\end{equation}
where the weight $\chi_j$ represents the projection of the nonlinear forcing onto the $j^{th}$ input mode
\begin{equation}\label{eq:weights}
\chi_j(k_x,k_z,\omega) = \displaystyle\int_{-1}^{1} \hat{\mathbf{f}}(k_x,k_z,\omega;y) \hat{\phi}_j^*(k_x,k_z,\omega;y) \mathrm{dy} 
\end{equation}
This particular decomposition was used by \citet{sharma2016low} to project ECS solutions to obtain low-dimensional representations. In what follows, we present an alternative formulation based on a more in-depth analysis of the nonlinear forcing term, and compare the merits of both techniques.

\subsection{Forcing decomposition}
We consider a Helmholtz decomposition of the nonlinear forcing term into the sum of an irrotational component and a solenoidal component,
\begin{equation}\label{eq:Helmholtz}
\mathbf{f} = \nabla \xi + \nabla \times \mathbf{\zeta} = \mathbf{f}_i + \mathbf{f}_s
\end{equation}
This decomposition requires a constraint to make it unique; here we choose $\nabla \cdot \mathbf{\zeta} = 0$ which manifests itself as enforcing homogeneous Dirichlet boundary conditions on ${f}_{v_s}$. This type of decomposition of the nonlinear term has previously been explored by \citet{wu1996reduced} and \citet{perot}. Substituting equation \ref{eq:Helmholtz} back into the NSE yields
\begin{equation}\label{eq:NSE_v2}
\frac{\partial \textbf{u}}{\partial t} - \mathbf{f}_s = -\nabla \tilde{p} + \frac{1}{Re_\tau} \nabla^2 \textbf{u}
\end{equation}
where we have absorbed the irrotational part of the nonlinear forcing into a modified pressure term $\tilde{p} = p - \xi$. As before, we can eliminate $\tilde{p}$ (and hence the irrotational forcing) and write this (Fourier-transformed) equation in terms of $\hat{v}$ and $\hat{\eta}$ as
\begin{equation}\label{eq:OS_SQ_v2}
-i\omega 
\left(
\begin{array}{c}
\hat{v} \\ \hat{\eta} 
\end{array}\right)
+\left(
\begin{array}{c c}
\Delta^{-1} & 0 \\ 0  & 1  \end{array} \right)
\left(
\begin{array}{c c}
\mathcal{L}_{OS} & 0 \\ ik_zU'  & \mathcal{L}_{SQ}
\end{array}\right)
\left(
\begin{array}{c}
\hat{v} \\ \hat{\eta} 
\end{array}\right)
= \left(
\begin{array}{c}
\hat{f}_{v_s} \\ \hat{f}_{\eta_s} 
\end{array}\right),
\end{equation}
where $\hat{f}_{\eta_s} \triangleq ik_z\hat{f}_{u_s} - ik_x\hat{f}_{w_s}$. Thus, the Helmholtz decomposition greatly simplifies the right-hand-side of equation \ref{eq:OS_SQ_v2} (in comparison to equation \ref{eq:OS_SQ}) and clearly reveals (upon elimination of the pressure) that only the solenoidal component of the nonlinear forcing is active. Furthermore, due to the decoupling of $\hat{v}$ from $\hat{\eta}$ (in a linear sense) and the fact that the forcing term in the $\hat{v}$ equation reduces to a single term, we can define the following direct input/output relationship
\begin{equation}\label{eq:v_fv}
\hat{v} = (-i\omega + \Delta^{-1}\mathcal{L}_{OS})^{-1}\hat{f}_{v_{s}},
\end{equation}
or equivalently
\begin{equation}\label{eq:v_fv_2}
\hat{v} = \mathcal{H}_v\Delta\hat{f}_{v_{s}},
\end{equation}
where
\begin{equation}
\mathcal{H}_v = (-i\omega\Delta + \mathcal{L}_{OS})^{-1}.
\end{equation}
The latter was found to be preferable in terms of numerical implementation. As before, we can perform an SVD of $\mathcal{H}_v$ and express $\hat{v}$ as
\begin{equation}\label{eq:v_separate}
\hat{v}(k_x,k_z,\omega;y) \approx \displaystyle\sum_{j=1}^{N} \sigma_{v_j}(k_x,k_z,\omega)\chi_{v_j}(k_x,k_z,\omega)\hat{\psi}_{v_j}(k_x,k_z,\omega;y) 
\end{equation}
where 
\begin{equation}
\chi_{v_j}(k_x,k_z,\omega) = \displaystyle\int_{-1}^{1} \Delta\hat{f}_{v_s}(k_x,k_z,\omega;y) \hat{\phi}_{v_j}^*(k_x,k_z,\omega;y) \mathrm{dy} 
\end{equation}
The equation that governs the vorticity can be written as
\begin{equation}
\hat{\eta} = \mathcal{H}_{\eta} \hat{f}_{\eta} -ik_z\mathcal{H}_{\eta}(U'\hat{v}) 
\end{equation}
where
\begin{equation}
\mathcal{H}_{\eta}(k_x,k_z,\omega) = (-i\omega + \mathcal{L}_{SQ})^{-1}.
\end{equation}
Thus we can express $\hat{\eta}$ as the sum of a component driven by the forcing $\hat{f}_{\eta}$ and a component driven by $\hat{v}$
\begin{equation}
\hat{\eta} = \hat{\eta}_f + \hat{\eta}_v, \hspace{5 mm} \hat{\eta}_f = \mathcal{H}_{\eta} \hat{f}_{\eta}, \hspace{5 mm} \hat{\eta}_v = -ik_z\mathcal{H}_{\eta}(U'\hat{v}).
\end{equation}
A low-order representation for $\hat{\eta}_v$ is simply obtained by substituting equation \ref{eq:v_separate} into $\hat{v}$. A SVD of $\mathcal{H}_{\eta}$ allows us to express $\hat{\eta}_f$ as
\begin{equation}\label{eq:eta_separate}
\hat{\eta}_f(k_x,k_z,\omega;y) \approx \displaystyle\sum_{j=1}^{N} \sigma_{\eta_j}(k_x,k_z,\omega)\chi_{\eta_j}(k_x,k_z,\omega)\hat{\psi}_{\eta_j}(k_x,k_z,\omega;y) 
\end{equation}
where 
\begin{equation}
\chi_{\eta_j}(k_x,k_z,\omega) = \displaystyle\int_{-1}^{1} \hat{f}_{\eta_s}(k_x,k_z,\omega;y) \hat{\phi}_{\eta_j}^*(k_x,k_z,\omega;y) \mathrm{dy} 
\end{equation}
The vorticity components $\hat{\eta}_v$ and $\hat{\eta}_f$ may be interpreted as nonlinear analogues to Orr-Sommerfeld/Squire modes. In fact, the left-hand-side of equation \ref{eq:OS_SQ_v2} looks identical to the Orr-Sommerfeld/Squire system for the analysis of the mean velocity. Thus the presence of the nonlinear term on the right-hand-side leads to a `forced' version of the traditional Orr-Sommerfeld/Squire system. Connections to Orr-Sommerfeld/Squire modes and their application to the study of turbulent flows are a topic of on-going work. We will now compare the advantages of this new approach based on separate decompositions for $\hat{v}$ and $\hat{\eta}$ with respect to the primitive variable approach.
% % % % % % % % % % % % % % % % % RESULTS % % % % % % % % % % % % % % % % % % % % % % % % % % % % % %
\section{Results}\label{sec:results}
\subsection{Low-dimensional representation}
For clarity, we will denote the results using the primitive variable formulation with solid symbols and the results using the separate decompositions with open symbols. We begin by presenting results for the primitive variable formulation in light of the notion of a forcing decomposition.  We demonstrate the efficacy of the singular modes obtained from the SVD of the resolvent operator as an efficient basis for capturing certain quantities of interest by projecting the P4 solutions onto these modes. We introduce the following scalar quantity $\gamma$ to characterize the percentage of a quantity of interest captured (in an integrated sense) as a function of the number of singular mode pairs ($N_p$)
\begin{equation}\label{eq:scalar}
\gamma  = \left[1 - \frac{|\langle q_{a},{r_{a}}^* \rangle-\langle q_e,{r_e}^*\rangle|}{\langle q_e,{r_e}^*\rangle}\right] \times 100
\end{equation}
where the inner product is defined as
\begin{equation}
\langle \hspace{1 mm}(\hspace{2 mm}),(\hspace{2 mm})^* \hspace{0.5 mm}\rangle = \displaystyle\int_{-1}^1 \displaystyle\int_{-\infty}^\infty \displaystyle\int_{-\infty}^\infty (\hspace{2 mm})(\hspace{2 mm})^* \mathrm{dk_x}\mathrm{dk_z}\mathrm{dy},
\end{equation}
the subscript `e' denotes the exact quantity computed directly from the simulation, the subscript `a' denotes the approximation based on using $N_p$ singular mode pairs, and $q$ and $r$ are flow quantities (such as velocity or forcing components). We present results in terms of singular pairs due to the imposed wall-normal symmetry mentioned earlier and the fact that the SVD yields symmetric/antisymmetric pairs. 

\begin{figure}
    \centering
    \begin{subfigure}[b]{0.4\textwidth}
        \includegraphics[width=\textwidth]{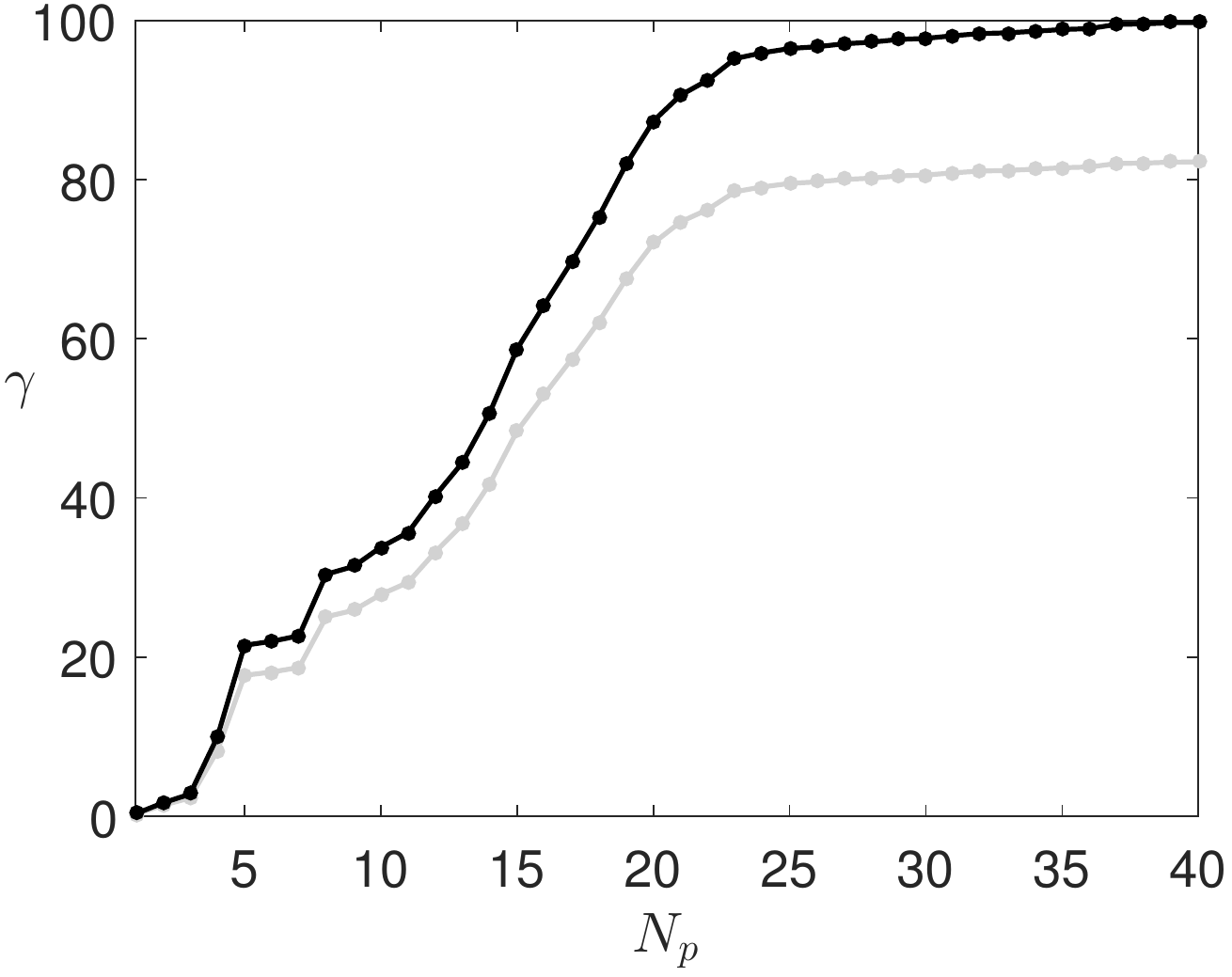}
        \caption{}
        \label{fig:FP4L}
    \end{subfigure}
    \begin{subfigure}[b]{0.4\textwidth}
        \includegraphics[width=\textwidth]{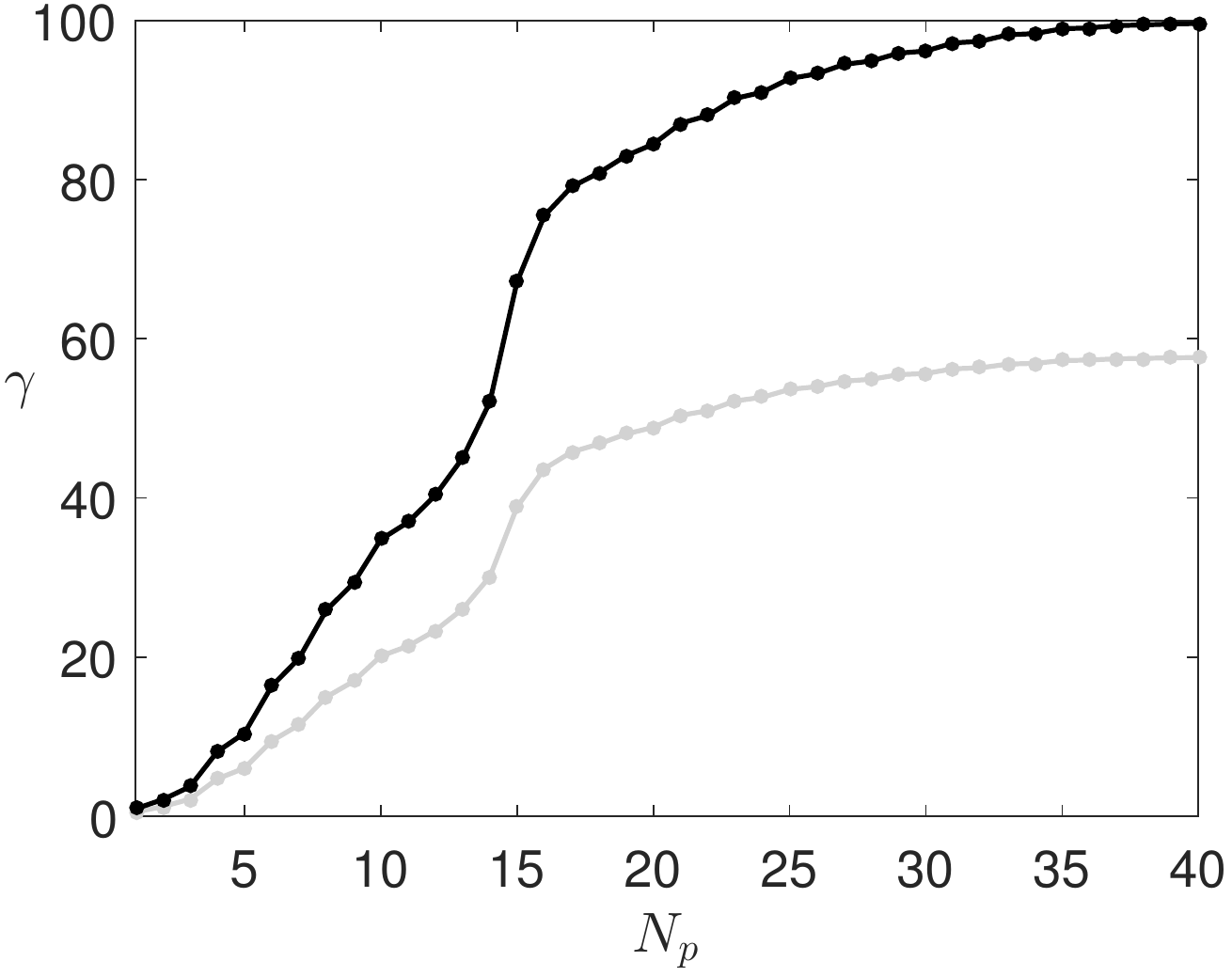}
        \caption{}
        \label{fig:FP4U}
    \end{subfigure}
    \caption{The percentage of the total nonlinear forcing captured (i.e. via projection of $-\langle \mathbf{u}' \cdot \nabla \mathbf{u}' \rangle_\mathbf{k}$ onto $\hat{\mathbf{\phi}}$) as a function of increasing number of singular mode pairs with respect to the full nonlinear forcing (gray) and solenoidal forcing (black) for the (a) P4L and (b) P4U solutions for the primitive formulation. Since the $\hat{\mathbf{\phi}}$'s span a divergence-free basis, they can only fully capture the solenoidal part of the forcing. The gap between the two curves represents the percentage of the total forcing that is irrotational.}\label{fig:forcing_normalization}
\end{figure}

Figure \ref{fig:forcing_normalization} shows the percentage of the full nonlinear forcing captured (see equation \ref{eq:weights}) for increasing number of singular modes for the P4L and P4U solutions. The gray curve is with respect to the full forcing, while the black curve uses the solenoidal forcing as the reference quantity defined in equation \ref{eq:scalar}. These curves illustrate that the projections (i.e. the basis spanned by $\hat{\mathbf{\phi}}$) will eventually capture all of the solenoidal forcing but cannot capture the remaining irrotational forcing, represented by the gap between the two curves. This is a consequence of removing pressure from the governing equations which constrains the input/response modes to be divergence free, as noted by \citet{luhar2014structure} and confirmed with the formal Helmholtz decomposition. \citet{sharma2016low} presented a similar figure and attributed the incomplete capturing of the full forcing to the selective filtering of the resolvent operator; we may now say more formally the irrotational forcing is in the null space of the operator defined in equation \ref{eq:PrimitiveResolvent}.

\begin{figure}
    \centering
    \begin{subfigure}[b]{0.3\textwidth}
        \includegraphics[width=\textwidth]{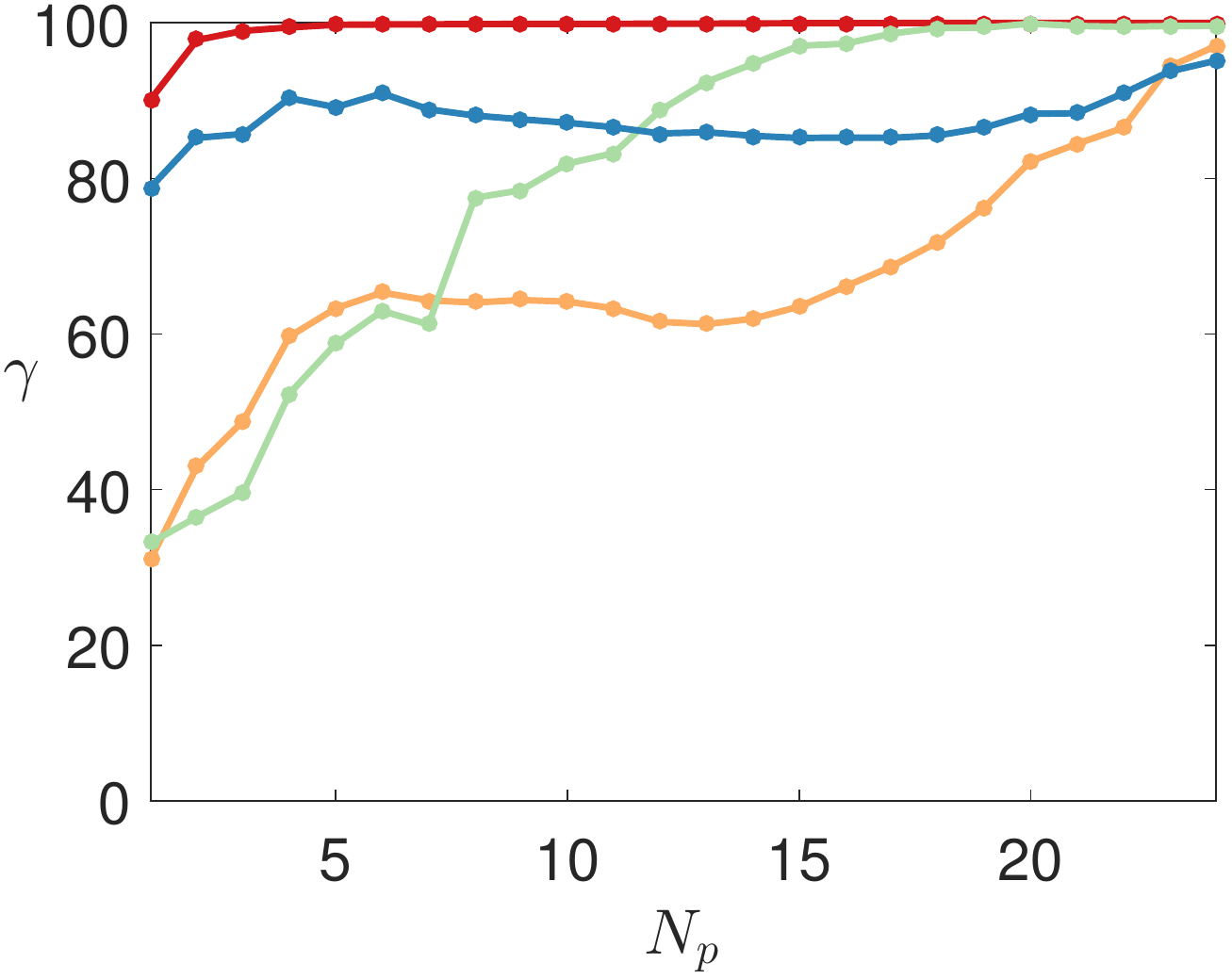}
        \caption*{(a)}
        \label{fig:UVWPL}
    \end{subfigure}
    \begin{subfigure}[b]{0.3\textwidth}
        \includegraphics[width=\textwidth]{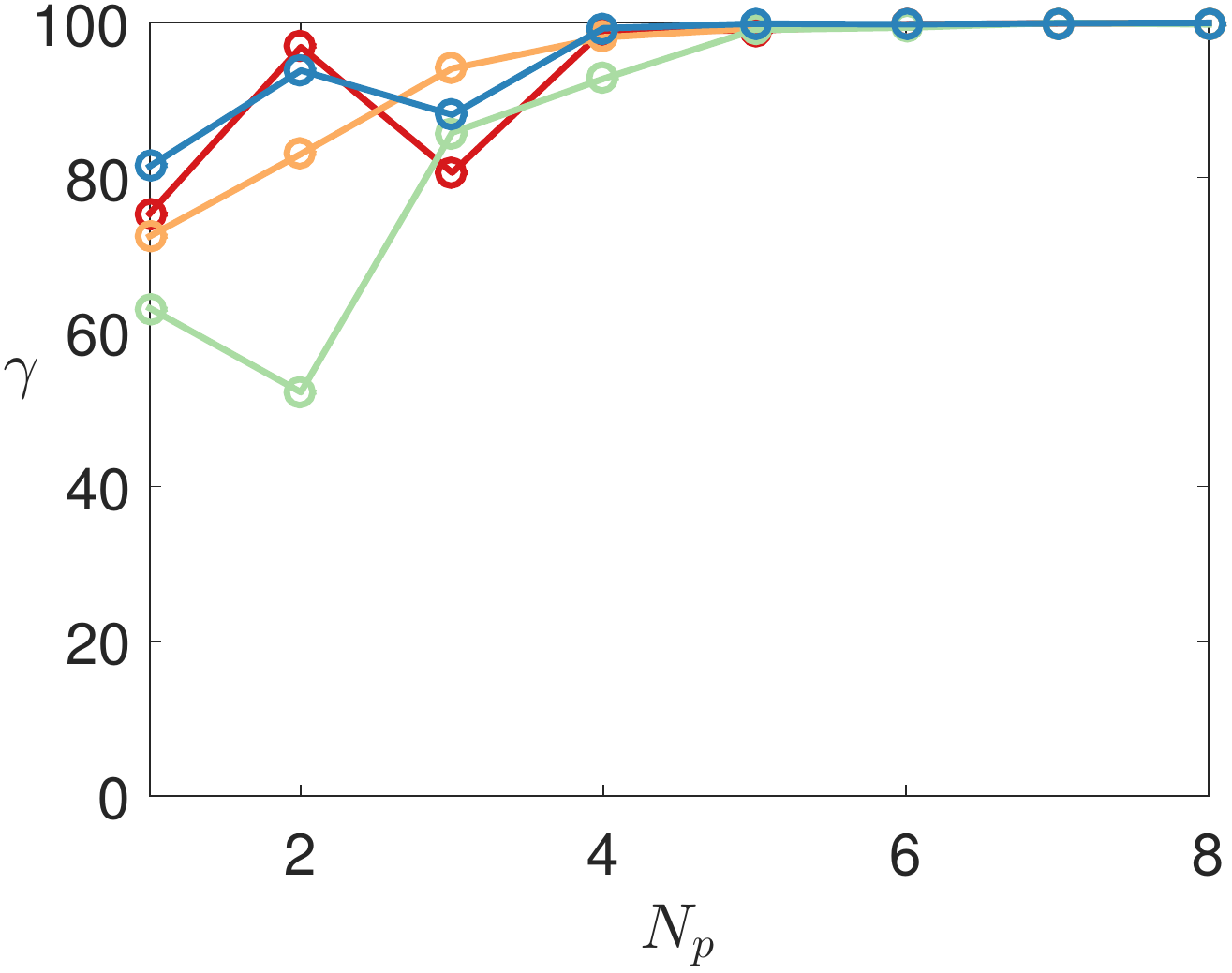}
        \caption*{(c)}
        \label{fig:UVWSL}
    \end{subfigure}
    \begin{subfigure}[b]{0.3\textwidth}
         \includegraphics[width=\textwidth]{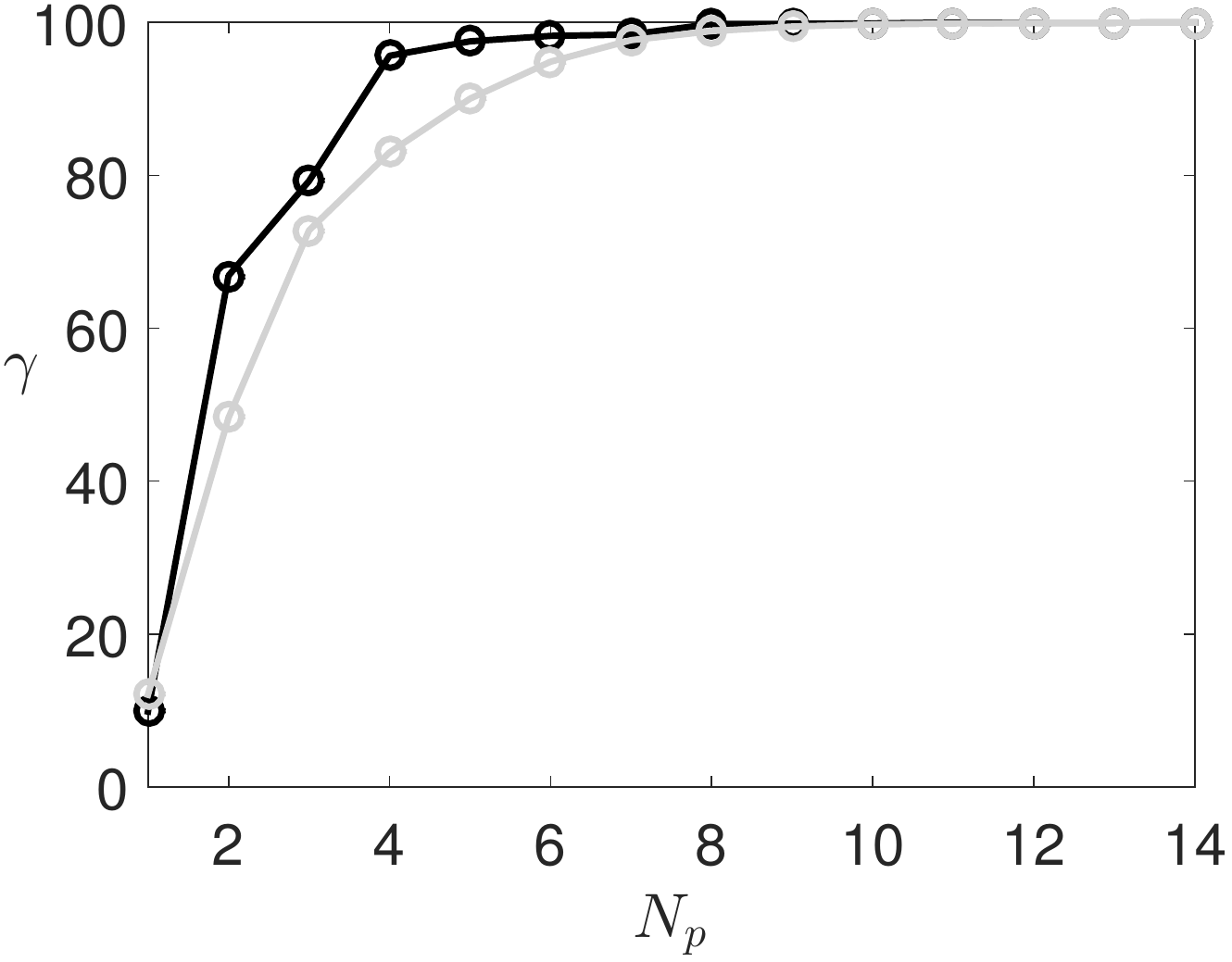}
         \caption*{(e)}
         \label{fig:FL}
    \end{subfigure}
    \begin{subfigure}[b]{0.3\textwidth}
         \includegraphics[width=\textwidth]{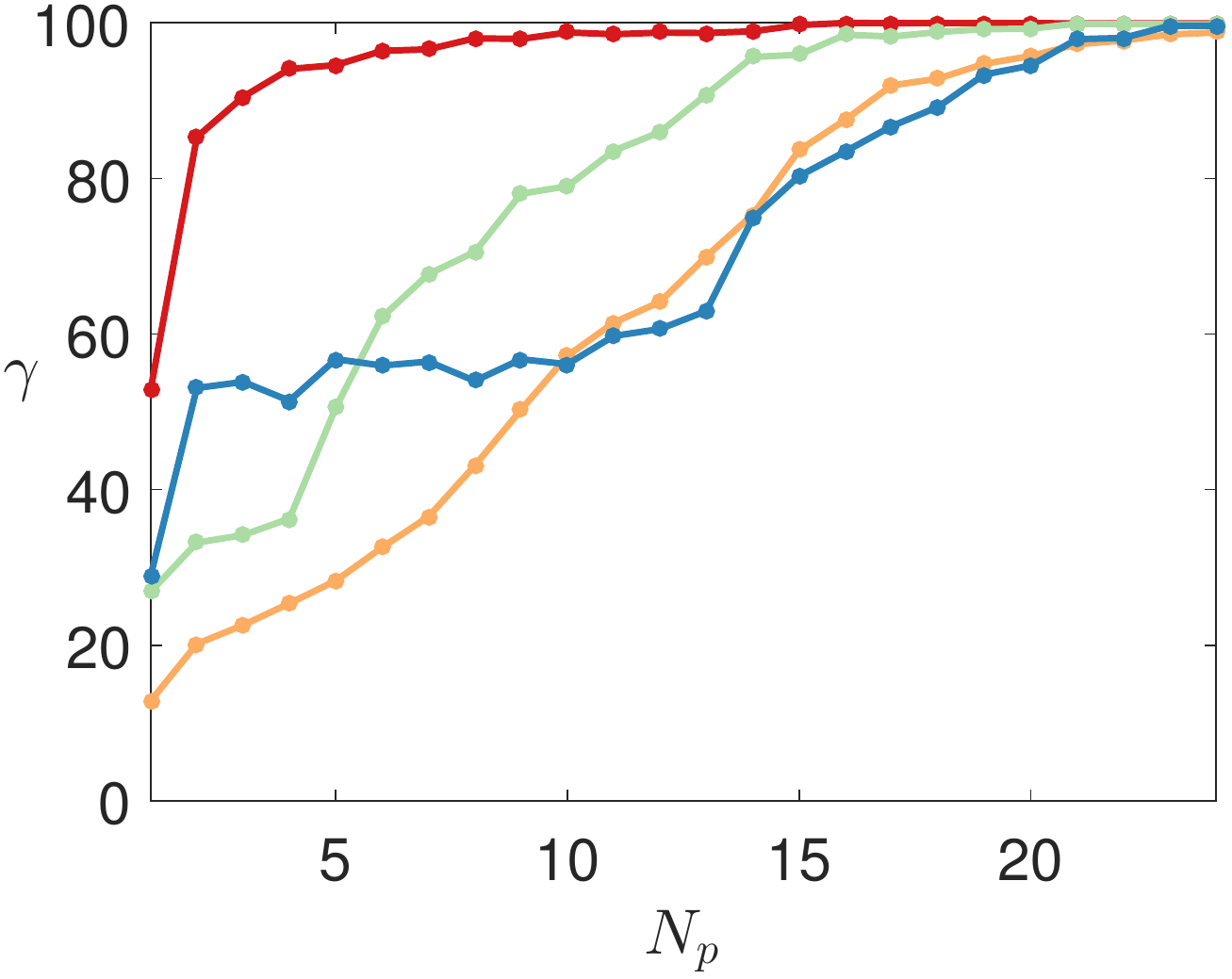}
         \caption*{(b)}
         \label{fig:UVWPU}
    \end{subfigure}
    \begin{subfigure}[b]{0.3\textwidth}
             \includegraphics[width=\textwidth]{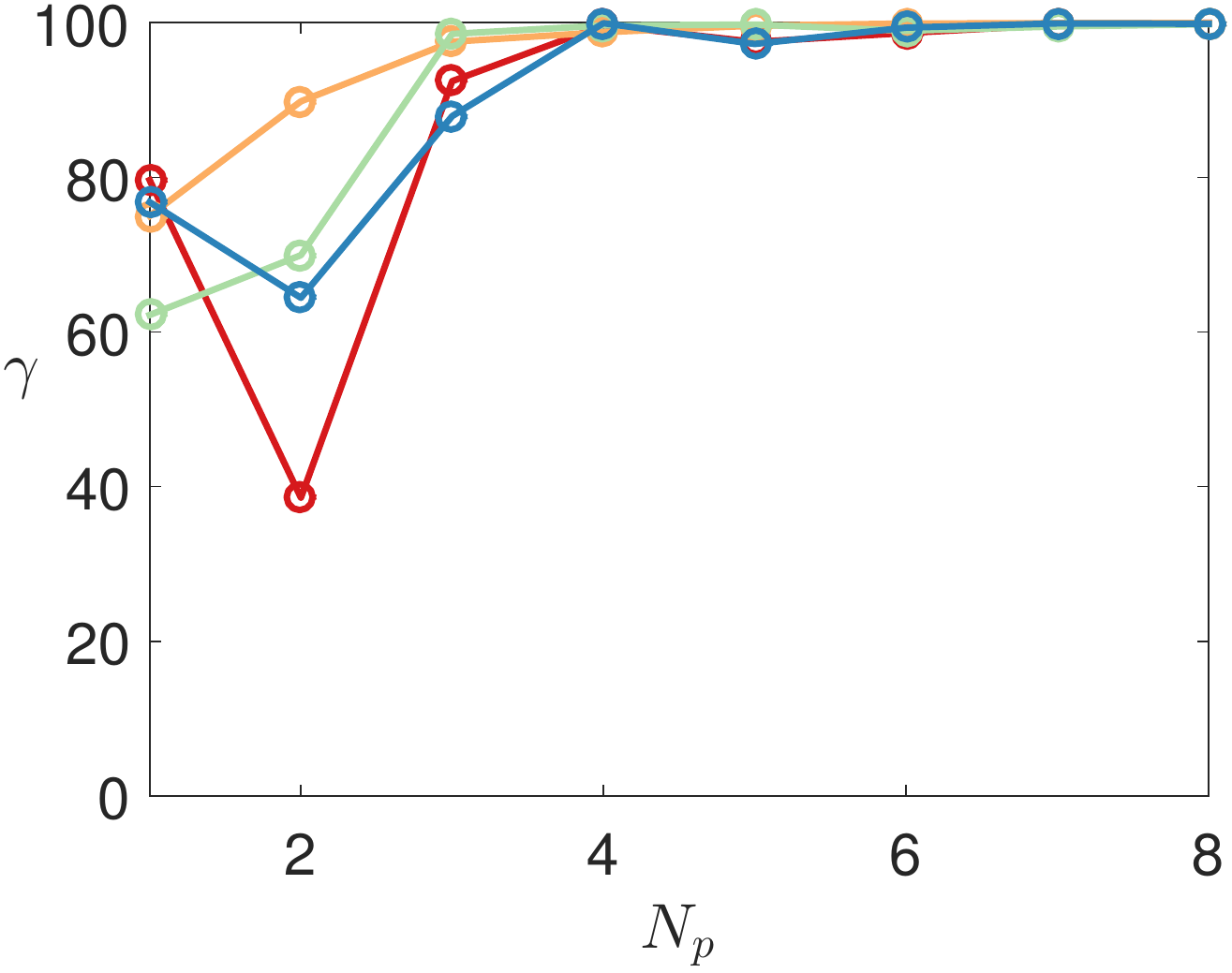}
             \caption*{(d)}
             \label{fig:UVWSU}
     \end{subfigure}
     \begin{subfigure}[b]{0.3\textwidth}
          \includegraphics[width=\textwidth]{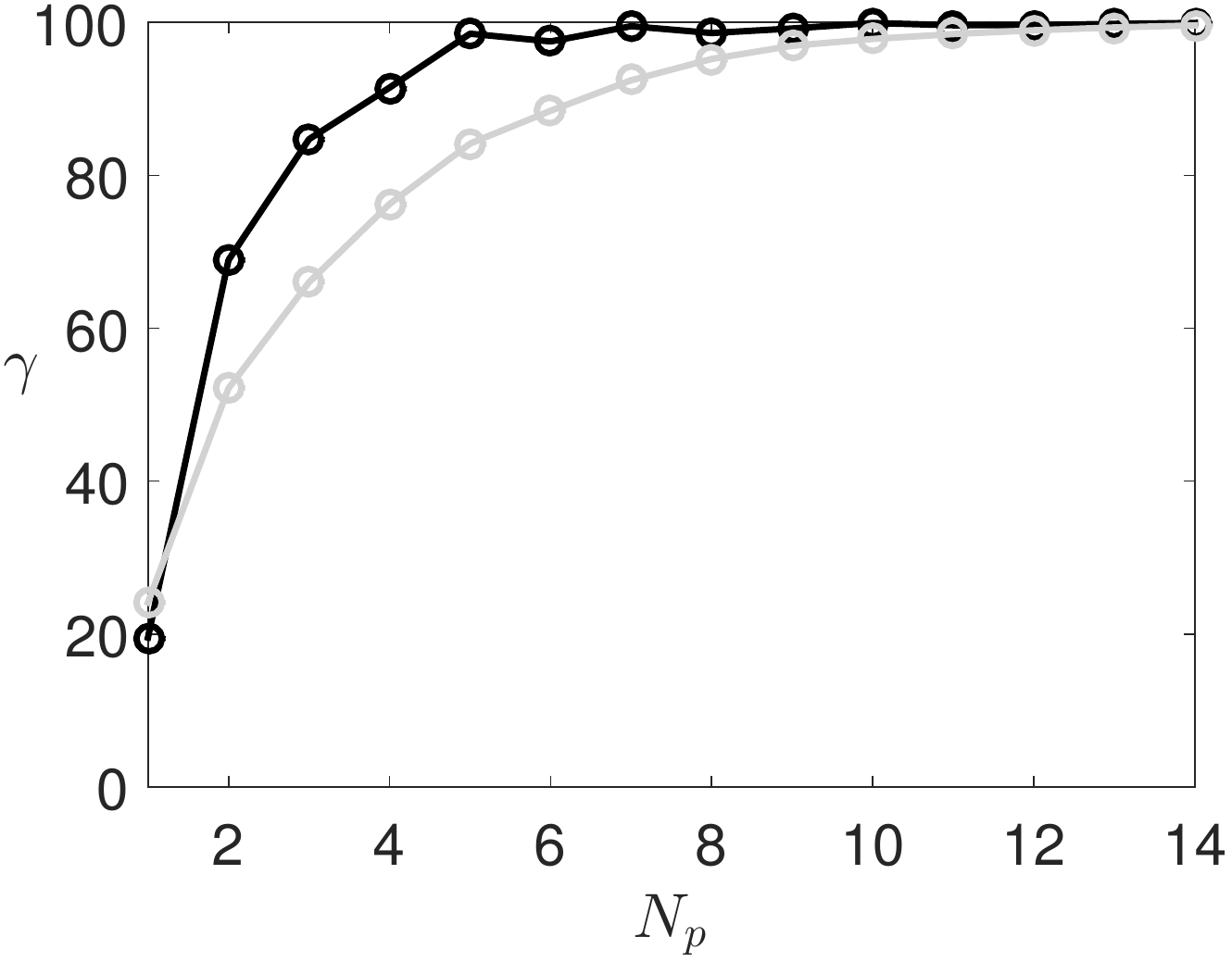}
          \caption*{(f)}
          \label{fig:FU}
     \end{subfigure}
    \caption{(a)-(d): The percentage of $\langle u^2 \rangle$ (red), $\langle v^2 \rangle$ (orange), $\langle w^2 \rangle$ (green), $\langle uv \rangle $ (blue) captured as a function of increasing number of singular mode pairs using a decomposition based on a primitive variable output (left column) and separate decompositions for $v$ and $\eta$ (middle column) for P4L (top row) and P4U (bottom row). (e)-(f): The percentage of $\langle {f_{v_s}}^2 \rangle$ (black) and $\langle {f_{\eta_s}}^2 \rangle$ (gray) captured as a function of increasing number of singular mode pairs for (e) P4L and (f) P4U. Note the change in scale of horizontal axis in each column.}\label{fig:decomposition_comparison}
\end{figure}

The true potential of the new decomposition technique is highlighted in figure \ref{fig:decomposition_comparison}. The left column presents results for capturing $\langle u^2 \rangle$, $\langle v^2 \rangle$, $\langle w^2 \rangle$, and $\langle uv \rangle$ using the primitive variable formulation while the middle column displays results based on using separate decompositions for $v$ and $\eta$ (and using equation \ref{eq:C_op} to calculate $u$ and $w$) for the P4L (top row) and P4U (bottom row) solutions. It should be noted similar results were presented for the primitive variable approach in \citet{sharma2016low}. We see that while the primitive variable approach captures $\langle u^2 \rangle$ quite well, the remaining components (notably $\langle uv \rangle$, which is required to sustain the mean velocity profile) converge much more slowly with increasing number of singular pairs. In contrast, the new approach captures all the components extremely efficiently, needing only roughly six singular pairs for both the P4L and P4U solutions. It is interesting that despite the increased spatial complexity of the P4U solution \citep[see][]{park2015exact}, it is still well approximated by the same number of singular pairs as the P4L solution.

The results in figure \ref{fig:decomposition_comparison} bring into question the choice of norm. As mentioned previously, the SVD dictates that the response modes in the primitive variable formulation are optimal with respect to a kinetic energy norm. However, in the case of the P4 solutions (and canonical turbulent flows of interest), the kinetic energy is often dominated by the streamwise velocity. As emphasized in figure \ref{fig:decomposition_comparison}, the separate decomposition approach does not suffer from this bias and equally captures all quantities very effectively.

The new decomposition technique appears to provide not only a very low-dimensional representation of the output, but also of the input forcing as well. Figure \ref{fig:decomposition_comparison} (right column) shows the percentage of the forcing components $\langle {f_{v_s}}^2 \rangle$ and $\langle {f_{\eta_s}}^2 \rangle$ captured for the P4L and P4U solutions. The forcing requires slightly more singular modes to be fully captured, particularly for the P4U solution; however, in comparison to the solenoidal forcing captured with the primitive variable approach (see figure \ref{fig:forcing_normalization}), this approach yields a much more compact representation.

\subsection{Forcing structure}

\begin{figure}
    \centering
    \begin{subfigure}[b]{0.33\textwidth}
        \includegraphics[width=\textwidth]{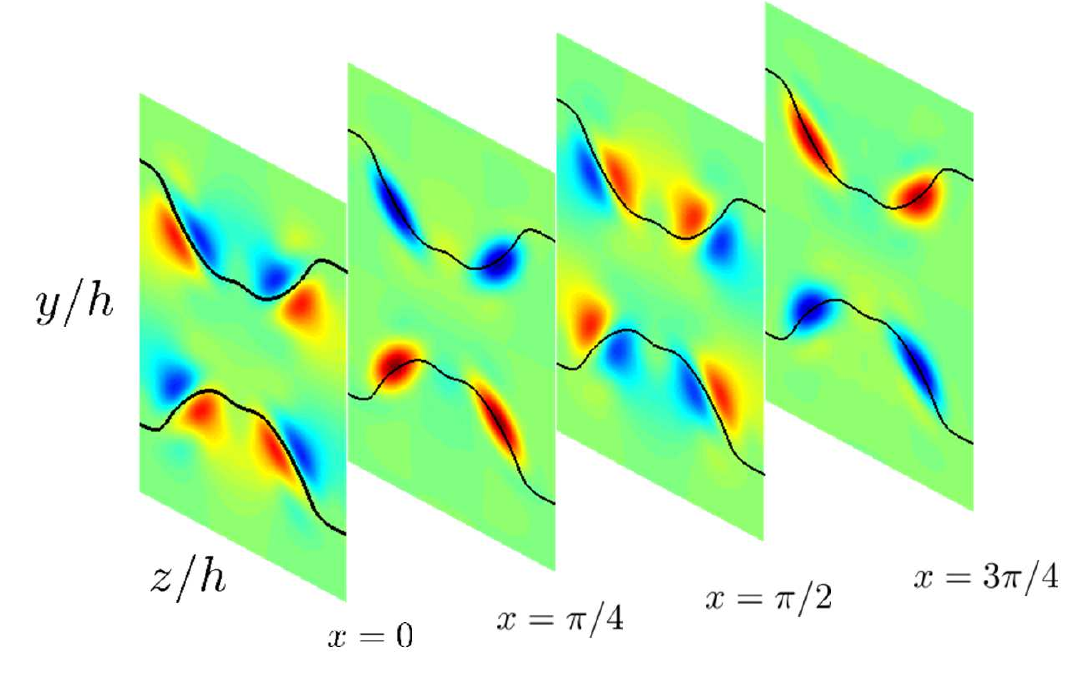}
        \caption*{(a)}
        \label{fig:vL}
    \end{subfigure}
    \begin{subfigure}[b]{0.33\textwidth}
        \includegraphics[width=\textwidth]{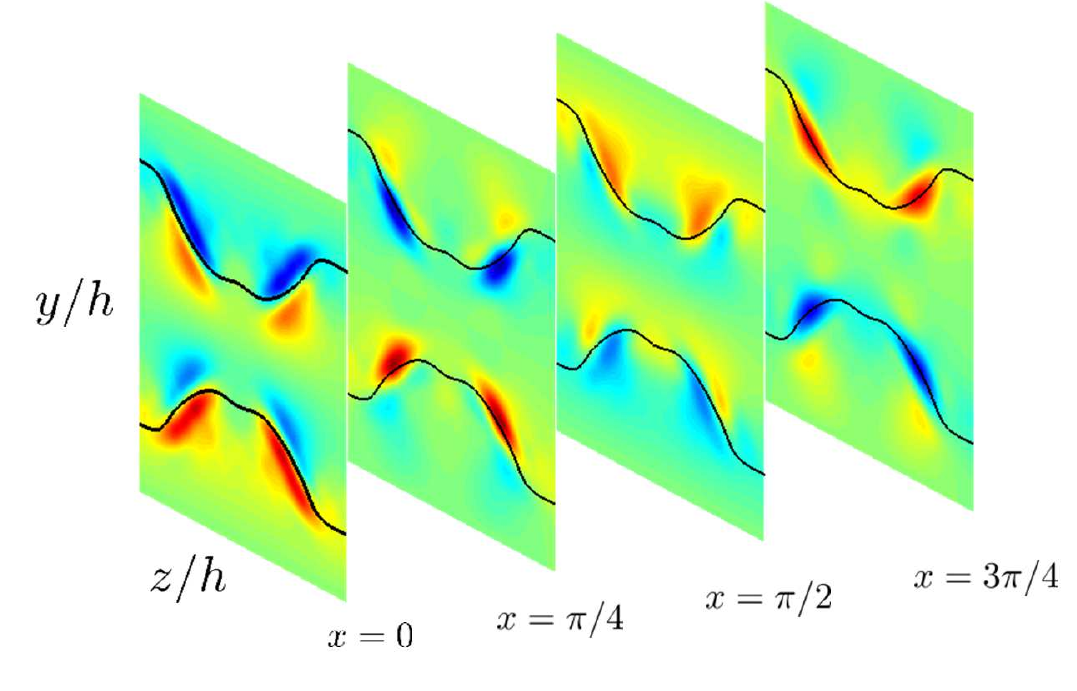}
        \caption*{(c)}
        \label{fig:fvL}
    \end{subfigure}
    \begin{subfigure}[b]{0.33\textwidth}
         \includegraphics[width=\textwidth]{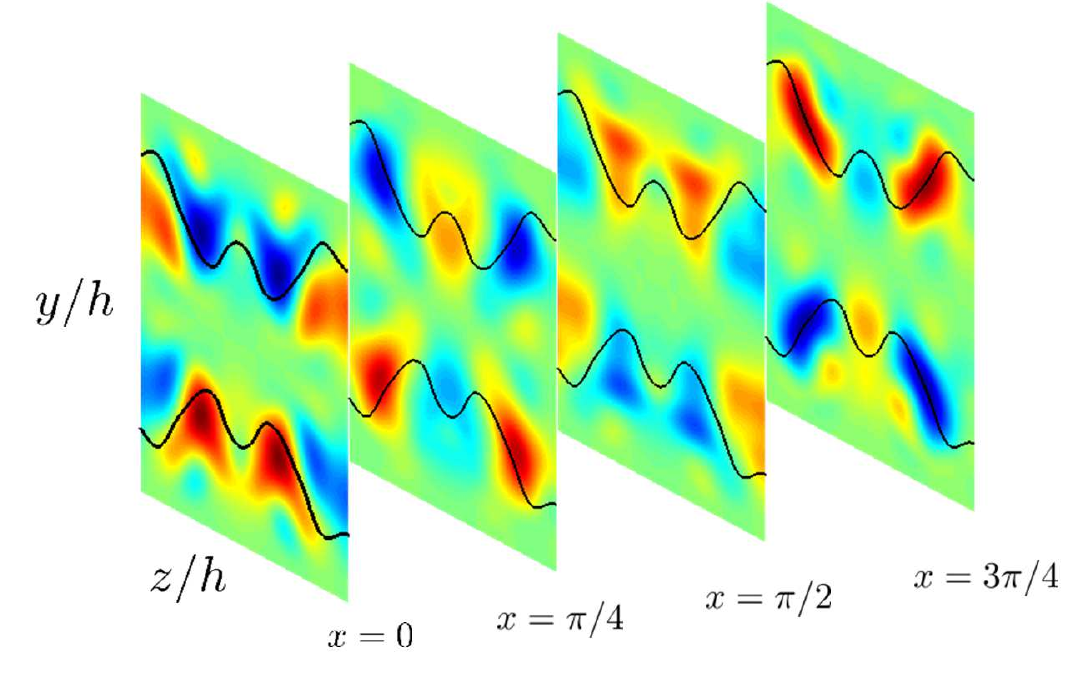}
         \caption*{(b)}
         \label{fig:vU}
    \end{subfigure}
    \begin{subfigure}[b]{0.33\textwidth}
         \includegraphics[width=\textwidth]{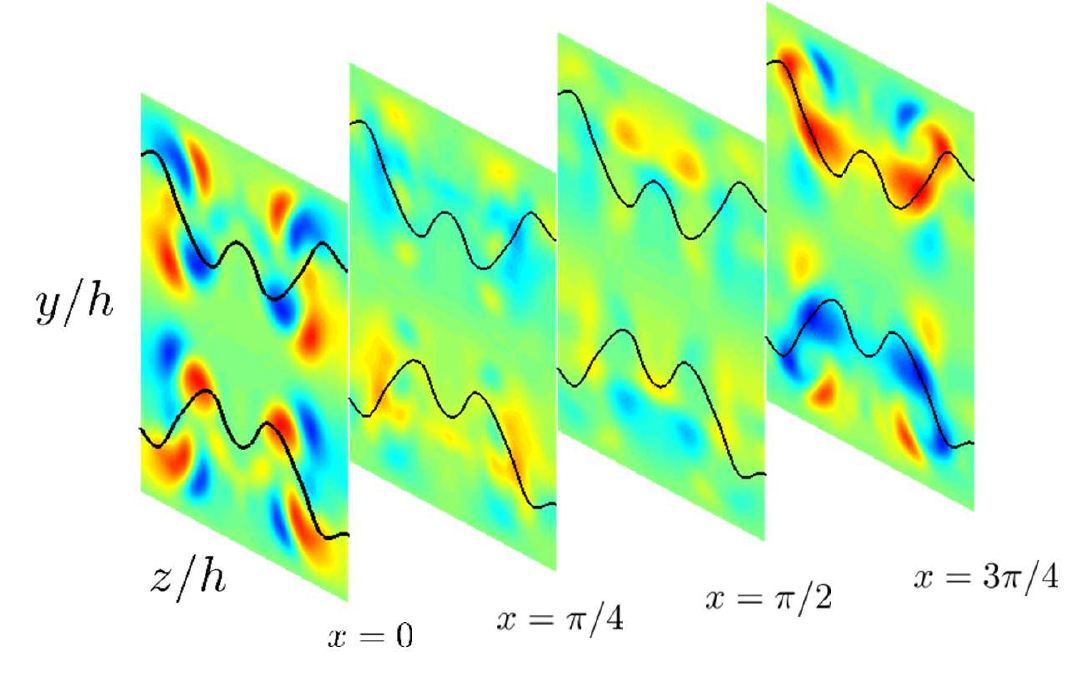}
         \caption*{(d)}
         \label{fig:fvU}
    \end{subfigure}
    \caption{Fluctuations of $v$ (a,b) and $f_{v_s}$ (c,d) with respect to a streamwise-averaged mean for P4L (top row) and P4U (bottom row) at four streamwise locations. The black lines denote the location of the critical layer defined as $\overline{U}(y,z) = c$.}\label{fig:v_fv_cl}
\end{figure}

We analyze the structure of the solenoidal forcing components, especially with respect to the critical layer. As has been studied previously for the P4 solutions due to the dominance of streamwise-constant modes, we consider a streamwise-averaged mean and analyze fluctuations about that mean (i.e. $q(x,y,z) = \overline{Q}(y,z)+q'(x,y,z)$ where q denotes a flow variable and the overbar represents a streamwise average). Consequently, we define the critical layer as the isocontour corresponding to $\overline{U}(y,z) = c$. However, it should be emphasized the preceding resolvent analysis considered fluctuations about a streamwise and spanwise-averaged mean. Figure \ref{fig:v_fv_cl} shows $v$ and $f_{v_s}$ fluctuations about a streamwise-averaged mean for the P4L and P4U solutions for multiple streamwise locations. It has been demonstrated \citep[see][]{park2015exact} that the velocity fluctuations for these solutions are localized about the critical layer (denoted by the black line), which is seen for the $v$ fluctuations. However, figure \ref{fig:v_fv_cl} also reveals that the forcing fluctuations are concentrated about the critical layer as seen for $f_{v_s}$, which is consistent with the results of \citet{hall2010streamwise}. We observed in equation \ref{eq:v_fv} there is direct input/output relationship between $v$ and $f_{v_s}$. The similarity in structure, particularly for the P4L solution, is striking. Though not reported here, $f_{\eta_s}$ also appeared to be localized about the critical layer. The overlap in physical space of the input and output and its role in the self-sustainment of these solutions is a topic of on-going work.
% % % % % % % % % % % % % % % % % CONCLUSION% % % % % % % % % % % % % % % % % % % % % % % % % % % % % %
\section{Conclusion}\label{sec:conclusions}
We demonstrated the utility of a Helmholtz decomposition of the nonlinear forcing term in revealing simplified input/output relationships in the resolvent analysis of the NSE. Notably, in the velocity-vorticity formulation, the decomposition highlighted the role of the solenoidal component of the forcing in the solution process. As opposed to the traditional resolvent formulation with primitive variable output, the new approach admitted separate decompositions for $v$ and $\eta$ and greatly reduced the number of singular modes needed to represent the velocity fields; while the primitive formulation required $N_p \approx 24$ per wavenumber pair, the new approach required only $N_p \approx 6$. In comparison to the number of wall-normal discretization points used in the simulation, this is a significant reduction in the number of degrees of freedom. Furthermore, the new approach also led to a compact representation of the forcing field, and we were able to demonstrate connections between the forcing structure and the critical layer.
\tikzstyle{block} = [draw, fill=gray!20, rectangle, 
    minimum height=3em, minimum width=6em, align=center, rounded corners, top color=white, bottom color=black!50]
\tikzstyle{input} = [coordinate]
\tikzstyle{output} = [coordinate]
\tikzstyle{pinstyle} = [pin edge={to-,thin,black}]
    \begin{figure}
    \centering
    \begin{tikzpicture}[auto, looseness=1, node distance=1.5cm,>=latex']
    \node [input, name=input] {};
    \node [block, right of=input, xshift = 2cm] (forcing) {$\mathbf{u}\cdot\nabla \mathbf{u}$};
    \node [output, right of=forcing, xshift = 2cm] (output) {};
    \node[block,below of=output,yshift = 0.2cm] (HD) {Helmholtz\\ decomposition};
    \node[output, below of=HD, yshift = 0.3cm] (output2) {};
    \node [block, below of=forcing, yshift = -1cm] (LD) {linear dynamics};
    \node[output,left of=LD, xshift = -2cm] (output3) {};
	\node[block, below of=input,yshift = 0.2cm] (Cblock) {C};
    % Once the nodes are placed, connecting them is easy. 
    \draw [draw,->] (input) -- node[pos=0.6] {$\mathbf{u}$}(forcing);
    \draw [-] (forcing) -- node {$\mathbf{f}$}(output);
    \draw [->] (output) -- (HD);
    \draw [-] (HD) -- (output2);
    \draw [->] (output2) -- node[align=center, above] {$\mathbf{f}_s$}(LD);
    \draw [-] (LD) -- node[align=center, above][pos=0.3] {$\left(\!\!\begin{array}{c} v \\ \eta \end{array}\!\!\right)$}(output3);
    \draw [->] (output3) -- (Cblock);
    \draw [-] (Cblock) -- (input);
    \end{tikzpicture}
    \caption{A block diagram representation of the NSE. The Helmholtz decomposition of the nonlinear forcing term illustrates how only the solenoidal component drives the velocity/vorticity fluctuations and thus is required in closing the loop.}
    \label{fig:BlockDiagram}
\end{figure}
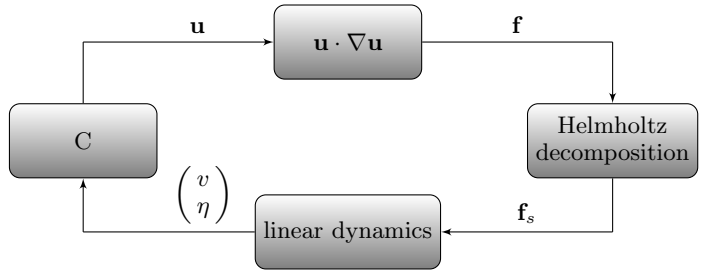

The notion of only the solenoidal forcing component being `active' in a sense, as illustrated in figure \ref{fig:BlockDiagram}, has interesting implications in terms of modeling Reynolds stresses as touched upon by \citet{jimenez}. Additionally, while this current analysis was performed in an incompressible setting (and leveraged this fact to make a clean distinction between solenoidal/irrotational forcing), it would be noteworthy to determine if an analogous decomposition of the nonlinear forcing could be made in a compressible context, as studied by \citet{towne2015stochastic}. The current analysis of ECS solutions may be extended to investigate not only travelling wave solutions, but periodic orbits as well as a step towards increased complexity. Ultimately, the hope is that a thorough knowledge of the modeling and behavior of the ECS solutions in the resolvent framework will inform the methodology used for the study of fully-turbulent flows.

\section*{Acknowledgments}
The authors gratefully acknowledge AFOSR (grant FA 9550-16-1-0361) for the financial support of this work and M.D. Graham and J.S. Park for providing the ECS data.

\bibliographystyle{jfm}
\bibliography{rapids_bibliography}

\end{document}